\documentclass[fleqn,usenatbib,usedcolumn]{mnras}
   \usepackage[british]{babel}             
   \usepackage{newtxtext}                  
   \usepackage[slantedGreek]{newtxmath}    
   \let\la=\lesssim 
   \usepackage[T1]{fontenc}                
   \usepackage{graphicx}                   
\title[ICME cross helicity]{Cross helicity of interplanetary coronal mass ejections at 1~au}

\author[S. W. Good et al.]{S. W. Good,$^{1}$\thanks{E-mail: simon.good@helsinki.fi}
L. M. Hatakka,$^{1}$
M. Ala-Lahti,$^{2,1}$
J. E. Soljento,$^{1}$
A. Osmane$^{1}$
and E. K. J. Kilpua$^{1}$
\\
$^{1}$Department of Physics, University of Helsinki, PO Box 64, 00014 Helsinki, Finland\\
$^{2}$Department of Climate and Space Sciences and Engineering, University of Michigan, 2455 Hayward St., Ann Arbor, MI 48109-2143, USA
}

\date{Accepted 2022 May 16. Received 2022 May 13; in original form 2022 April 1}

\pubyear{2022}

\begin{document}
\label{firstpage}
\pagerange{\pageref{firstpage}--\pageref{lastpage}}
\maketitle

\begin{abstract}

Interplanetary coronal mass ejections (ICMEs) contain magnetic field and velocity fluctuations across a wide range of scales. These fluctuations may be interpreted as Alfv\'enic wave packets propagating parallel or anti-parallel to the background magnetic field, with the difference in power between counter-propagating fluxes quantified by the cross helicity. We have determined the cross helicity of inertial range fluctuations at $10^{-3}-10^{-2}$~Hz in 226 ICME flux ropes and 176 ICME sheaths observed by the \textit{Wind} spacecraft at 1~au during 1995--2015. The flux ropes and sheaths had mean, normalised cross helicities of 0.18 and 0.24, respectively, with positive values here indicating net anti-sunward fluxes. While still tipped towards the anti-sunward direction on average, fluxes in ICMEs tend to be more balanced than in the solar wind at 1~au, where the mean cross helicity is larger. Superposed epoch profiles show cross helicity falling sharply in the sheath and reaching a minimum inside the flux rope near the leading edge. More imbalanced, solar wind-like cross helicity was found towards the trailing edge and laterally further from the rope axis. The dependence of cross helicity on flux rope orientation and the presence of an upstream shock are considered. Potential origins of the low cross helicity in ICMEs at 1~au include balanced driving of the closed-loop flux rope at the Sun and ICME-solar wind interactions in interplanetary space. We propose that low cross helicity of fluctuations is added to the standard list of ICME signatures.

\end{abstract}




\section{Introduction}

Interplanetary coronal mass ejections \citep[ICMEs; e.g.,][]{Kilpua17a} are vast, transient eruptions of plasma and magnetic field from the Sun with properties that are distinct from the solar wind. Signatures of ICMEs may variously include low plasma-$\beta$, a smooth rotation of the magnetic field at large scales, and magnetic fluctuations with small amplitudes \citep[e.g.,][]{Burlaga91,Cane03}. ICMEs with the first two of these signatures are classified as magnetic clouds. ICMEs displaying the second signature may be modelled as flux ropes, with twisted field lines in the ICME volume that wind around a common central axis \citep[e.g.,][]{Riley04,AlHaddad13}. A spacecraft encountering an ICME flux rope will observe a background magnetic field vector that monotonically rotates by up to 180$^{\circ}$ during the rope passage time. ICMEs are of significant interest as space weather drivers, given that their flux rope structures are a major source of southward magnetic field arriving at Earth \citep[e.g.,][]{Wilson87,Zhang88,Kilpua17b}. Sheath intervals of compressed solar wind ahead of ICMEs can also be highly geoeffective \citep[e.g.,][]{Tsurutani88,Lugaz16,Kilpua17a,Kilpua19}.

Like the solar wind, ICME plasma displays broadband magnetic field and velocity fluctuations with power spectra that take power law forms \citep{Leamon98,Liu06,Hamilton08,Borovsky19,Sorriso21}. In the solar wind, most power is contained in Alfv\'enically polarised fluctuations, with typically a few per cent of total power in other, compressive modes \citep[e.g.,][]{Chen16}. It has long been known that Alfv\'enic fluctuations with an anti-sunward sense of propagation predominate over sunward fluctuations in the solar wind \citep{Belcher71}, consistent with the fluctuations having mostly a coronal origin. This anti-sunward predominance is present in fluctuations at low frequencies with an $f^{-1}$ power spectrum and, to a somewhat lesser extent, in turbulent fluctuations at higher frequencies \citep[e.g.,][]{Chen13}. The predominance is also typically greater in fast wind than in slow, and declines with distance from the Sun \citep[e.g.,][]{Luttrell87,Marsch90,Bavassano00}. The amplitudes of sunward and anti-sunward fluctuations were found by \citet{Borovsky12} and \citet{Borovsky19} to be more equally balanced in ICME plasma at 1~au than in other solar wind types, apart from intervals containing sector reversals in the interplanetary magnetic field. A similar balance was also found by \citet{Good20} in a magnetic cloud observed at 0.25~au by \textit{Parker Solar Probe} \citep[PSP;][]{Fox16}. Given that the non-linear interaction of counter-propagating Alfv\'enic fluctuations is the source of Alfv\'enic turbulence in the solar wind, the degree of balance or imbalance has a significant effect on the turbulence properties that are observed \citep[e.g.,][and references therein]{Schekochihin20}.

Imbalance may be quantified in terms of the normalised cross helicity, $\sigma_\textrm{c}$, which gives the difference in power between Alfv\'enic fluctuations propagating parallel and anti-parallel to a background magnetic field. We here present analysis of $\sigma_\textrm{c}$ for a large number of ICME flux ropes and their upstream sheath regions observed by the \textit{Wind} spacecraft at 1~au over a 20-year period. Fluctuations across a decade of frequencies typically falling within the inertial range of magnetohydrodynamic (MHD) turbulence are investigated, at scales smaller than that of the background flux rope structure. We find that average values of $\sigma_\textrm{c}$ in the rope and sheath intervals are more balanced (i.e., $\sigma_\textrm{c}$ is closer to zero) than in the solar wind generally, consistent with previous results. Via superposed epoch analysis, systematic variations in $\sigma_\textrm{c}$ through the sheath--ICME complex along the heliocentric radial direction have been identified. The dependencies of $\sigma_\textrm{c}$ on the crossing distance of the spacecraft from the flux rope axis, on the presence or absence of an upstream shock bounding the sheath, and on the flux rope orientation, are also considered.

Cross helicity may be used to define Alfv\'enicity \citep[e.g.,][]{Stansby19}. This definition equates high Alfv\'enicity with the presence of unidirectional Alfv\'en waves, since such waves have well correlated or anti-correlated magnetic field and velocity fluctuations, and cross helicity is ultimately a measure of this correlation. In this work, cross helicity is interpreted in terms of Alfv\'enic balance or imbalance, and the normalised residual energy, $\sigma_\textrm{r}$, is used to define and measure Alfv\'enicity. Residual energy, which gives the difference in power between velocity and magnetic field fluctuations, is zero for idealised Alfv\'en waves but is generally observed to be negative (indicating an excess of magnetic fluctuation energy) in the solar wind inertial range \citep[e.g.,][]{Chen13}. Note that all mentions of cross helicity and residual energy in this work refer to normalised quantities.

In Section~\ref{sec:data}, the spacecraft dataset is described and key parameters are defined. The properties of an example ICME flux rope and sheath interval are described in detail in Section~\ref{sec:example}, followed by analysis of mean properties across all intervals in Section~\ref{sec:mean}. In Section~\ref{sec:ICME_dependence}, the superposed epoch analysis is presented, which considers sub-structuring within the intervals. Possible origins of the low cross helicity in ICMEs and their sheaths are discussed in Section~\ref{sec:disc}.

\section{Spacecraft Data}
\label{sec:data}

ICME flux ropes catalogued by \citet[][]{Nieves18} using \textit{Wind} spacecraft data have been analysed. The catalogue\footnote{https://wind.nasa.gov/ICMEindex.php} lists 272 intervals with flux rope rotations of the background magnetic field vector that were observed during 1995--2015; intervals in which 5 per cent or more of magnetic field or plasma data were missing have been excluded, leaving 226 intervals included in the present analysis. A total of 176 sheath intervals listed in the catalogue that meet the 5 per cent data gap threshold have also been analysed, of which 97 were associated with shocks.

Magnetic field data from MFI \citep{Lepping95} and ion moments from 3DP/PESA-L \citep{Lin95} on board \textit{Wind} have been used. Data were typically at 3~s resolution, with linear interpolation applied to close any small data gaps. Measurements of the magnetic field, $\textit{\textbf{B}}$, proton velocity, $\textit{\textbf{v}}$, and proton number density, $n_\textrm{p}$, have been used to determine the Elsasser variables, $\textit{\textbf{z}}^{\pm}=\textit{\textbf{v}}\pm\textit{\textbf{b}}$, where $\textit{\textbf{b}}=\textit{\textbf{B}}/\sqrt{\mu_0\rho}$ and $\rho$ is the ion mass density. It has been assumed that 4\% of solar wind ions are alpha particles and the rest protons by number density such that $\rho=7m n_\textrm{p}/6$, where $m_\textrm{p}$ is the unit proton mass. Fluctuations in $\textit{\textbf{z}}^{+}$ and $\textit{\textbf{z}}^{-}$ correspond to Alfv\'enic wave packets propagating anti-parallel and parallel to the background magnetic field, respectively.

The trace spectral densities of $\textit{\textbf{v}}$, $\textit{\textbf{b}}$ and $\textit{\textbf{z}}^{\pm}$, denoted by $E_v$, $E_b$ and $E_{\pm}$, respectively, have been calculated with a Morlet wavelet technique \citep{Torrence98} and used to obtain the normalised cross helicity,
\begin{equation}
\sigma_\textrm{c}=\frac{E_{+}-E_{-}}{E_{+}+E_{-}}
\end{equation}
and normalised residual energy,
\begin{equation}
\sigma_\textrm{r}=\frac{E_v-E_b}{E_v+E_b}.
\end{equation}
A similar wavelet method has been applied previously by \citet{Chen13} to calculate $\sigma_\textrm{c}$ and $\sigma_\textrm{r}$. Values of $\sigma_\textrm{c}$ and $\sigma_\textrm{r}$ are limited by definition to the range $[-1, 1]$.

\section{Analysis}
\label{sec:analysis}

\subsection{An example event}
\label{sec:example}

\begin{figure*}
	\includegraphics[width=\textwidth]{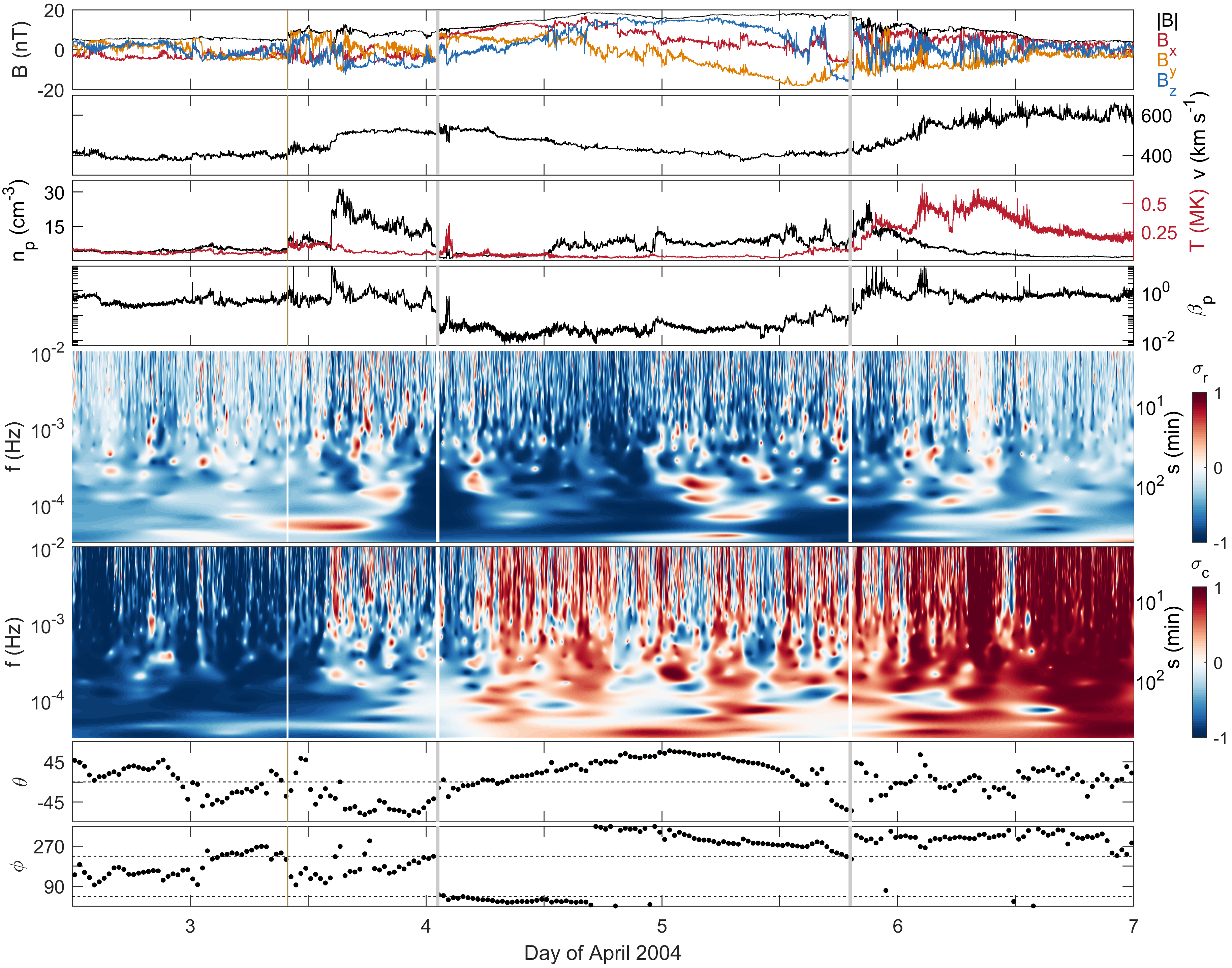}
    \caption{An example ICME. From top to bottom, the panels show: the magnetic field in GSE coordinates; bulk proton speed; proton density and temperature; proton plasma-$\beta$; normalised residual energy; normalised cross helicity; and 30-min averages of the magnetic field latitude and longitude angles in GSE coordinates. The ICME boundaries and upstream shock are marked by grey and gold vertical lines, respectively.}
    \label{fig:event_example}
\end{figure*}

Figure~\ref{fig:event_example} shows \textit{Wind} observations for one of the ICMEs analysed. The flux rope interval, bounded by vertical grey lines, displays the characteristic rotation of the $\textit{\textbf{B}}$ vector through a large angle. This rotation is evident in the bottom two panels, which show the latitude angle of $\textit{\textbf{B}}$ relative to the ecliptic plane, $\theta$, and the anti-clockwise angle between the GSE-\textit{x} (radially sunward) direction and the projection of $\textit{\textbf{B}}$ onto the ecliptic, $\phi$. The $\theta$ and $\phi$ angles are averaged to a resolution of 30~min; we suggest that variations at this timescale are predominately due to the global structure of the flux rope. At this scale, $\textit{\textbf{B}}$ rotated from approximately solar east to north to west through the interval, consistent with the passage of a tube of twisted field lines wound around a central axis with a northward-pointing inclination to the ecliptic. The ordered $\textit{\textbf{B}}$ rotation, enhanced $|\textit{\textbf{B}}|$ and low proton plasma-$\beta$ ($\lesssim 0.1$) indicate that the ICME was a magnetic cloud. Horizontal dotted lines in the $\phi$ panel denote the nominal sector boundaries of the interplanetary magnetic field assuming a Parker spiral angle of $45^{\circ}$, with $\phi$ values between (outside) the lines indicating field in the away (toward) sector. The sheath interval was bounded by a shock\footnote{http://www.ipshocks.fi}, marked by the vertical gold line, and the leading edge of the flux rope. A second shock was present within the sheath.

Wavelet spectrograms of $\sigma_\textrm{r}$ and $\sigma_\textrm{c}$ in the frequency range $3.2\times10^{-5}-10^{-2}$~Hz are displayed in Figure~\ref{fig:event_example}. Spectrograms of time intervals longer than those shown in Figure~\ref{fig:event_example} were obtained, such that the intervals which are shown in the figure fall entirely within the cone of influence. Upstream of the shock, slow wind with strongly negative $\sigma_\textrm{c}$ was present. Since the mean field was mostly in the away sector in this upstream region, negative $\sigma_\textrm{c}$ (i.e., greater power in $\textit{\textbf{z}}^{-}$ than $\textit{\textbf{z}}^{+}$ fluctuations) here indicates the prevalence of fluctuations with an anti-sunward sense of propagation in the plasma frame. Anti-sunward fluctuations were likewise prevalent in the faster solar wind trailing the ICME during 6~April, where $\sigma_\textrm{c}$ was strongly positive and the mean field was in the toward sector. In both the upstream and downstream intervals, $|\sigma_\textrm{r}|$ was low (indicative of fairly Alfv\'enic wind) and almost entirely negative.

The flux rope interval displayed a patchy mix of positive and negative $\sigma_\textrm{c}$, with a globally averaged value close to zero. This local patchiness in $\sigma_\textrm{c}$ is a fundamental feature of balanced turbulence as predicted by models \citep{Perez09}. Values of $\sigma_\textrm{r}$ were lower in the flux rope than in the ambient solar wind, particularly at the lower end of the frequency range. As in the upstream wind, $\sigma_\textrm{c}$ was globally negative in the sheath, but with significant positive patches that increased the mean value. Globally, there was little difference between $\sigma_\textrm{r}$ in the sheath and ambient wind.

In the following, we restrict our analysis to the frequency range $10^{-3}-10^{-2}$~Hz (equivalent to wave periods $1.67-16.7$~min). The inertial range of MHD turbulence usually encompasses these frequencies at 1~au \citep[e.g.,][]{Kiyani15}. At much larger scales (e.g., frequencies $\lesssim 10^{-4}$~Hz, periods $\gtrsim 2.8$~hr), a significant fraction of fluctuation power in an ICME may be due to the variation of the flux rope B-field. Rather than this background, mean-field structure, we focus on smaller scale fluctuations present within the structure. At $10^{-3}-10^{-2}$~Hz, the ICME shown in Figure~\ref{fig:event_example} had globally averaged $\{\sigma_\textrm{c},\sigma_\textrm{r}\}$ values of $\{0.10,-0.47\}$ in the flux rope and $\{-0.38,-0.26\}$ in the sheath. The sample points used to determine these averages were equally spaced across the logarithmic frequency range.

\subsection{Mean values}
\label{sec:mean}

\begin{figure}
	\includegraphics[width=0.9\columnwidth]{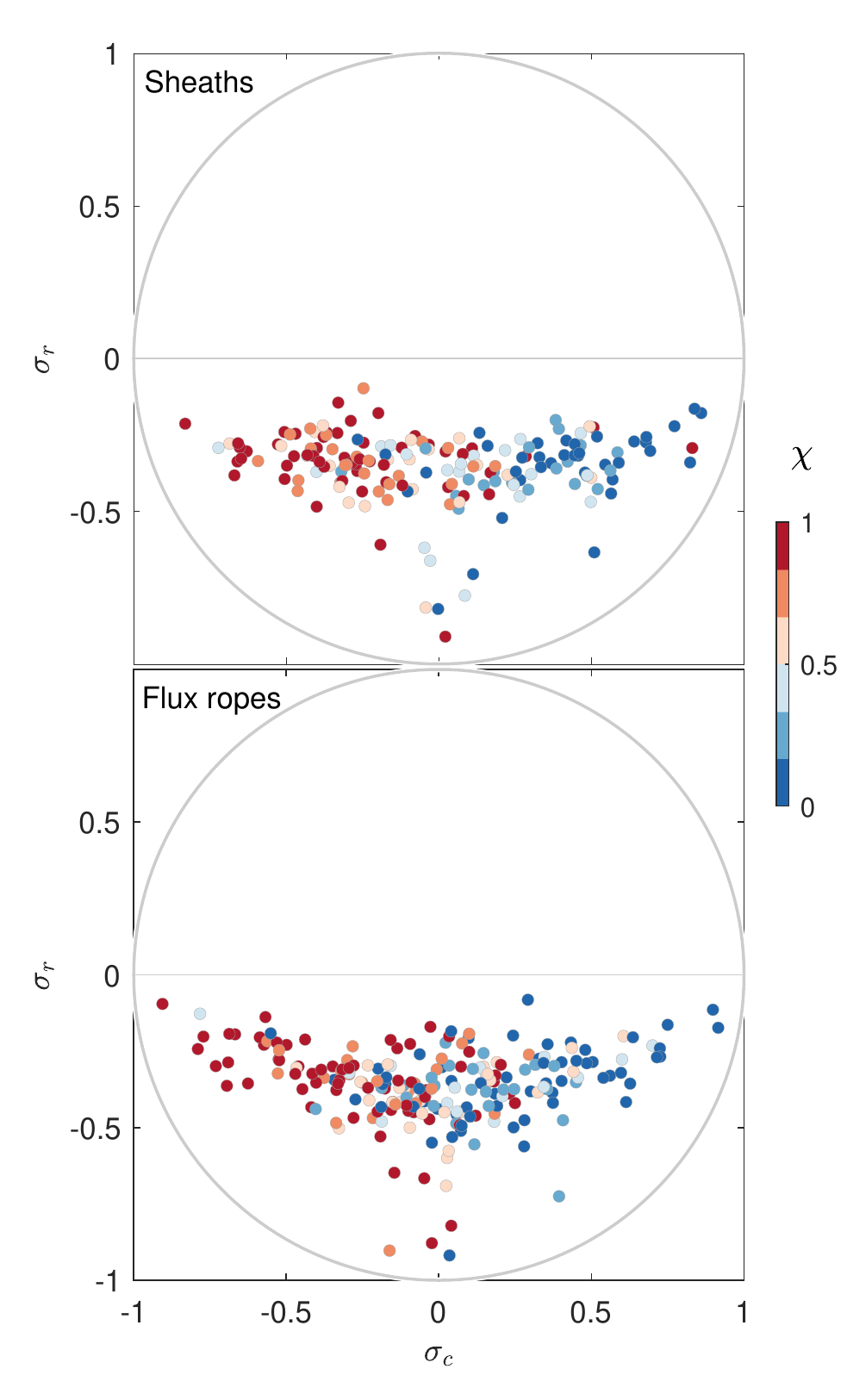}
    \caption{Globally averaged values of $\sigma_\textrm{r}$ versus $\sigma_\textrm{c}$ across the upstream sheath regions (top panel) and flux rope intervals (bottom panel). The $\chi$ parameter gives the fraction of the interval in the away sector.}
    \label{fig:averages_figure}
\end{figure}

Figure~\ref{fig:averages_figure} shows the globally averaged values of $\sigma_\textrm{r}$ versus $\sigma_\textrm{c}$ at $10^{-3}-10^{-2}$~Hz for the 226 flux rope and 176 sheath intervals analysed. Values are constrained by definition to fall within the circle described by $\sigma_\textrm{r}^2+\sigma_\textrm{c}^2=1$. The mean of the globally averaged $\sigma_\textrm{r}$ values for the flux rope intervals was $-0.36$, similar to the mean of $-0.35$ found for the sheath intervals. None of the flux rope or sheath intervals had a globally positive value of $\sigma_\textrm{r}$. 

Points in Figure~\ref{fig:averages_figure} are colour-coded according to the fraction of data in the corresponding interval that was in the away sector, $\chi$. Red ($\chi>0.5$) and blue ($\chi<0.5$) points indicate intervals that were primarily in the away and toward sectors, respectively. For the sheath intervals, the sector structure is defined relative to the Parker spiral as described in Section~\ref{sec:example}. For the flux ropes, which do not have a Parker spiral field, we define a sector structure based on the distributions of GSE $\phi$ angles observed within ICMEs obtained by \citet{Borovsky10a}. In Figure~3 of \citet{Borovsky10a}, broad peaks at around $120^{\circ}$ and less distinctly at $300^{\circ}$ can be seen in the distributions for three different ICME catalogues. These two angles are used as the midpoints of the away and toward sectors for the flux rope intervals (cf. $135^{\circ}$ and $315^{\circ}$ for the equivalent Parker spiral sectors).

The anti-correlation of $\sigma_\textrm{c}$ and $\chi$ that can be seen in Figure~\ref{fig:averages_figure} is consistent with the prevalence of anti-sunward fluctuations: when the mean field is in the away (toward) sector, $\sigma_\textrm{c}$ tends to be negative (positive) and fluctuations in the anti-sunward $\textit{\textbf{z}}^{-}$ ($\textit{\textbf{z}}^{+}$) mode are dominant. A meaningful average value of $\sigma_\textrm{c}$ across all intervals is obtained by first rectifying the magnetic field, so that positive $\sigma_\textrm{c}$ is defined as the anti-sunward flux regardless of the interval sector. This rectification is achieved by reversing the mean field direction for the away sector intervals, equivalent to reversing the sign of $\sigma_\textrm{c}$ for the points with $\chi>0.5$ in Figure~\ref{fig:averages_figure}. The mean value of rectified cross helicity, $\sigma_\textrm{c}^*$, was $0.18$ for the flux rope intervals and $0.24$ for the sheaths, in both cases closer to zero than the average solar wind value of $0.40$ found by \citet{Chen13}. In practice, the mean $\sigma_\textrm{c}^*$ value for the flux ropes is the same for sector boundaries defined relative to the Borovsky distributions and the Parker spiral, although the $\chi$ values for some individual intervals are sensitive to the choice of sector boundary definition.

A low average value of $|\sigma_\textrm{c}|$ in a particular interval may arise simply if the interval has a mixed sector structure. When excluding intervals with $1/6<\chi<5/6$ in Figure~\ref{fig:averages_figure} (i.e., paler-shaded points), which have a more mixed structure, the mean values of $\sigma_\textrm{c}^*$ become $0.24$ for the flux ropes and $0.31$ for the sheaths, higher than the all-interval means but still lower than the solar wind average. This suggests that the low cross helicity found in this analysis is only partially due to a mixing of sectors, and that it is a more intrinsic property of the fluctuations.

\begin{table}
	\centering
	\caption{Mean interval-averaged values of $\sigma_\textrm{r}$ and $\sigma_\textrm{c}^*$ in ICME sheaths and flux ropes. Standard deviations (SD) are also listed.}
	\label{tab:averages}
	\begin{tabular}{lcccc}
		\hline
		 & \multicolumn{2}{c}{Residual energy} & \multicolumn{2}{c}{Cross helicity} \\
		 & $\sigma_\textrm{r}$ & SD & $\sigma_\textrm{c}^*$ & SD\\
		\hline
		Sheaths & $-0.35$ & 0.12 & 0.24 & 0.30 \\
		Flux ropes & $-0.36$ & 0.13 & 0.18 & 0.29 \\
		\hline
	\end{tabular}
\end{table}

\subsection{Dependence on ICME parameters}
\label{sec:ICME_dependence}

\begin{figure}
\centering

	\includegraphics[width=0.85\columnwidth]{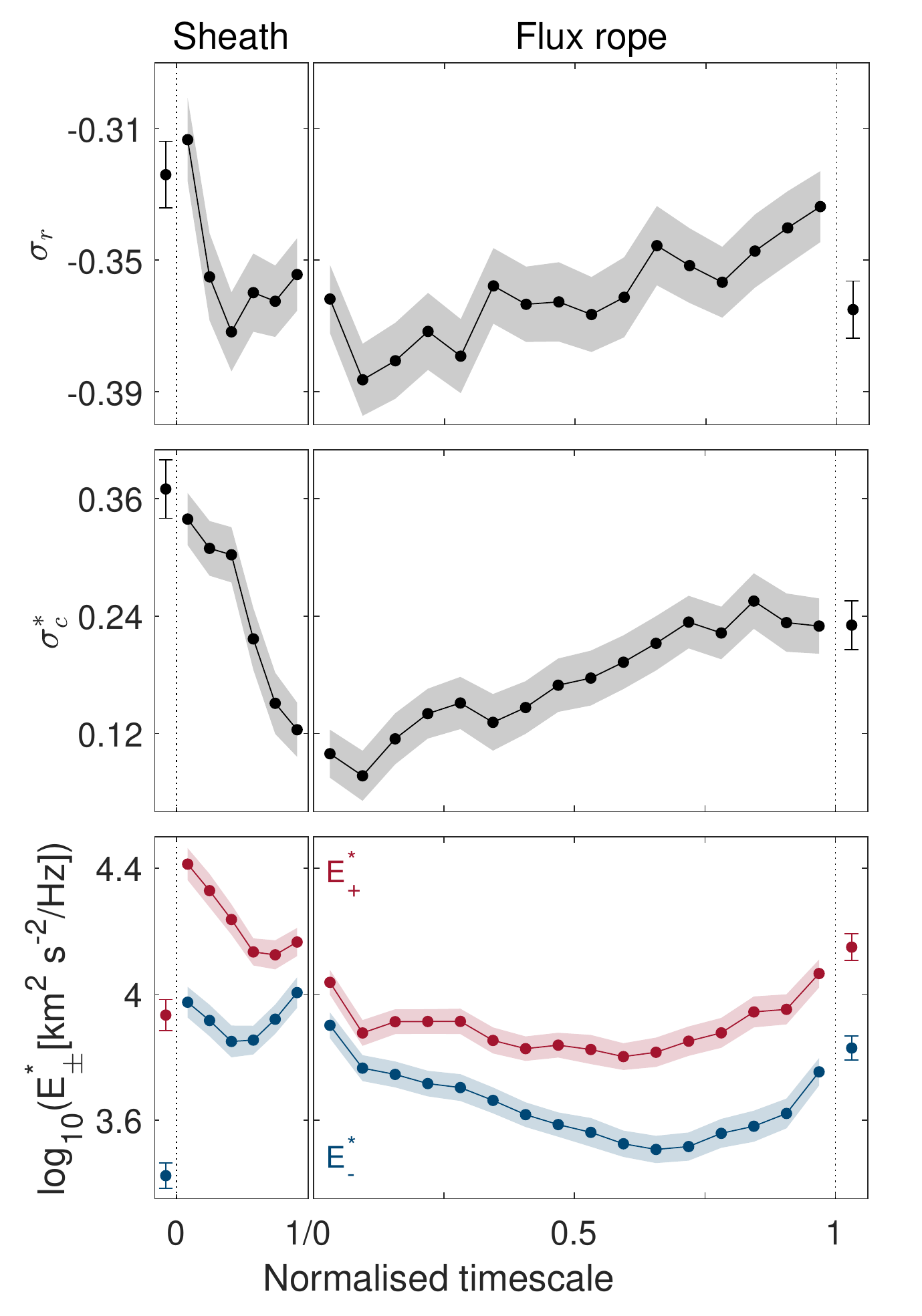}
    \caption{Superposed epoch profiles of $\sigma_\textrm{r}$, $\sigma_\textrm{c}^*$ and $E^*_\pm$ for sheaths (left panels) and flux ropes (right panels), made by binning all events. Dashed lines indicate the start and end of the ICME profile. Single bins spanning 6~hr of solar wind before and after the ICMEs are also shown. Shading and error bars indicate the standard error of values in each bin.}
    \label{fig:SEA_figure1}
\end{figure}

\begin{figure}
	\includegraphics[width=0.99\columnwidth]{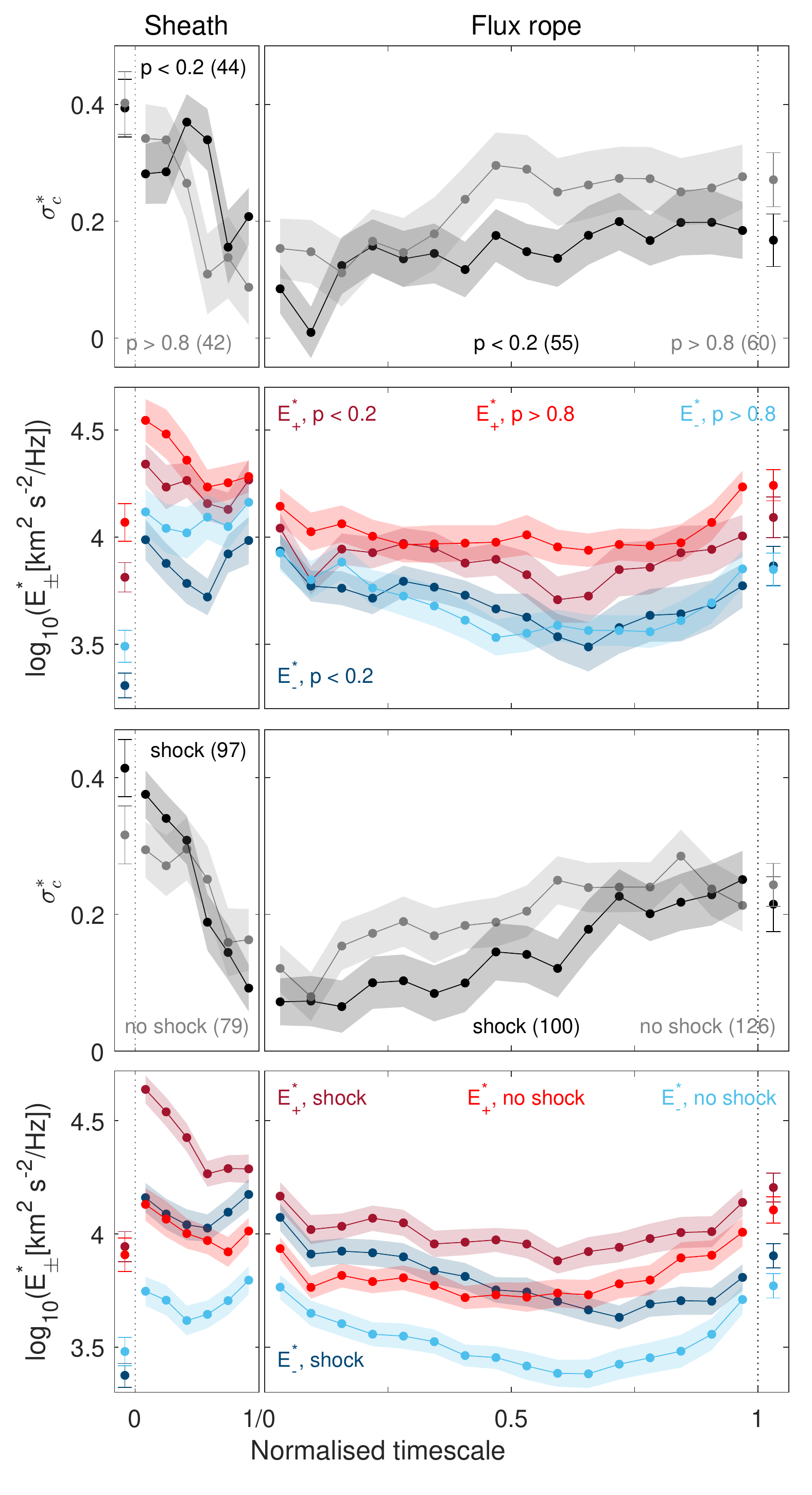}
    \caption{Superposed epoch profiles of $\sigma_\textrm{c}^*$ and $E^*_\pm$ binned according to high and low impact parameter, $p$, and to the presence or absence of an upstream shock, in a similar format to Figure~\ref{fig:SEA_figure1}. The number of events contributing to each profile is given in parentheses in the $\sigma_\textrm{c}^*$ panels.}
    \label{fig:SEA_figure2}
\end{figure}

ICMEs are often observed by spacecraft as highly structured intervals with systematic variations in properties. Variations may, for example, be seen from the front to the back of an ICME flux rope as it passes over an observing spacecraft \citep[e.g.,][]{Lynch03,Masias16,Rodriguez16}. Properties also vary depending on flux rope orientation, and on which region of the rope's 3D structure that happens to be observed. Sheaths likewise show systematic variations in properties from their leading edges to the interface with the driver material \citep{Kilpua19,Salman20,Kilpua21}, and also in the lateral direction \citep{Ala-Lahti20}. Spatial variations in $\sigma_\textrm{c}^*$ and $E^*_\pm$ from the front to back of the sheath-flux rope complex, and the dependence of these variations on a range of global ICME parameters, are considered. Here, $E^*_+$ and $E^*_-$ represent power in fluctuations with an anti-sunward and sunward sense of propagation in the plasma frame, respectively (cf. unrectified $E_\pm$).

Flux rope fitting parameters obtained by \citet{Nieves19} have been used to categorise the 226 flux rope intervals. The fits were performed with the circular-cylindrical model derived by \citet{Nieves16}, which characterises flux ropes as axially symmetric cylinders of twisted magnetic field. Fitting of the model to the observed $\textit{\textbf{B}}$ timeseries with a chi-square minimisation gives various global parameters of the flux rope; of relevance here are the latitude, $\theta_0$, and longitude, $\phi_0$, of the flux rope axis direction in GSE coordinates, the flux rope handedness, $H$, which may be either left (-1) or right (+1) handed, and the impact parameter, $p=|y_0/R|$, which gives the distance of closest approach made by the spacecraft to the flux rope axis, $y_0$, as a fraction of the rope radius, $R$.

\subsubsection{All-event profiles}
\label{sec:SEA1}

Figure~\ref{fig:SEA_figure1} shows superposed epoch profiles of $\sigma_\textrm{r}$, $\sigma_\textrm{c}^*$ and $E^*_\pm$ for the flux ropes and sheaths. These profiles were produced by normalising the time span of each rope and sheath interval, and then dividing each into 16 and 6 sub-intervals, respectively. The average value of each parameter in each sub-interval was determined from the wavelet spectrograms, then averaged across all events by sub-interval number to give the profiles shown. Sub-intervals in a particular event with 5 per cent or more of data missing were excluded. Rectification to obtain $\sigma_\textrm{c}^*$ and $E^*_\pm$ was performed with the $\chi$ value calculated in each sub-interval, using the method described in Section~\ref{sec:mean}. This procedure was applied to the rope and sheath intervals separately. The mean durations of the flux rope and sheath sub-intervals were approximately equal, at 1.5~hr and 1.6~hr, respectively. Single points for intervals spanning 6~hr immediately preceding the sheath and following the rope are also shown. The profiles in Figure~\ref{fig:SEA_figure1} represent cuts in the radial heliocentric direction through the sheath-flux rope complex. The leading and trailing edges of each profile have normalised timescale values of `0' and `1', respectively; the trailing edge of the sheath (at `1' on the sheath profile $x$-axis) is equivalent to the leading edge of the flux rope (at `0' on the rope profile $x$-axis). 

The $\sigma_\textrm{r}$ profile displays a sharp drop followed by a plateau at lower values in the sheath, and an uneven rise from the leading to trailing edge of the flux rope. The more Alfv\'enic $\sigma_\textrm{r}$ value of the first data point in the sheath profile ($-0.31$) is attributed to fluctuations generated immediately downstream of shocks: the profile including only sheaths with shocks (not shown) displays this more Alfv\'enic downstream region, while the equivalent profile for sheaths without shocks is flat throughout, at $\sigma_\textrm{r}\sim -0.36$. The spread in $\sigma_\textrm{r}$ values across the profile is relatively narrow.

The $\sigma_\textrm{c}^*$ profile shows, from front to back, a slight decrease from the upstream solar wind in the front half of the sheath, a sharp drop towards the sheath rear, a minimum within the flux rope close to the sheath-flux rope interface, and a gradual rise towards the flux rope trailing edge. The average $\sigma_\textrm{r}$ and $\sigma_\textrm{c}^*$ values across the sheath and rope profiles are consistent with the values in Table~\ref{tab:averages}, as expected. While remaining at low values, the average $\sigma_\textrm{c}^*$ is in all locations positive, indicating that the balance is tipped towards the anti-sunward direction throughout the ICME.

It is instructive to consider variations in $E^*_+$ and $E^*_-$ individually, recalling that $\sigma_\textrm{c}^*$ is the normalised difference between these two quantities. The $E^*_\pm$ profiles in Figure~\ref{fig:SEA_figure1} show the typical dominance of $E^*_+$ over $E^*_-$ in the solar wind upstream of the sheath. From upstream wind to sheath, there is a sharp increase in $E^*_+$ and $E^*_-$, with both then showing a central dip and rise towards the sheath rear. It can be seen that the drop in $\sigma_\textrm{c}^*$ at the sheath rear is due to a relatively greater enhancement in $E^*_-$ than $E^*_+$. In the front two-thirds of the flux rope, the flat $E^*_+$ combined with a falling $E^*_-$ causes the rising $\sigma_\textrm{c}^*$ noted previously. Both $E^*_+$ and $E^*_-$ gradually rise towards the back of the rope.

\subsubsection{Impact parameter and shocks}
\label{sec:SEA2}

The first and second rows in Figure \ref{fig:SEA_figure2} show profiles separated according to whether the impact parameter through the flux rope was high ($p>0.8$) or low ($p<0.2$). Lower $\sigma_\textrm{c}^*$ can generally be seen across the rope at low $p$, when the spacecraft trajectory came closer to the central axis, while higher, more solar wind-like $\sigma_\textrm{c}^*$ tends to be found at high $p$, further from the central axis. This $\sigma_\textrm{c}^*$ vs. $p$ trend is primarily due to higher $E^*_+$ at large $p$ within the rope, although the differences in $E^*_\pm$ between high and low $p$ encounters is not large. There are weak signatures of correlated W-shaped profiles in $E^*_+$ and $E^*_-$ across the rope at low $p$.

The bottom two rows of Figure~\ref{fig:SEA_figure2} display profiles for sheaths and ropes sorted according to whether or not an upstream shock was present. It can be seen that $E^*_\pm$ are significantly higher for the shock-associated profiles, to the extent that $E^*_-$ in the shock-associated profiles is equal to or greater than $E^*_+$ in the non-shock profiles through the sheath and front half of the rope. The resulting differences in $\sigma_\textrm{c}^*$ are less significant, however, with $\sigma_\textrm{c}^*$ being somewhat lower in the shock-associated profile through the rope, and about the same in both profiles through the sheath.

\subsubsection{Flux rope structure}
\label{sec:SEA3}

\begin{figure}
	\includegraphics[width=0.99\columnwidth]{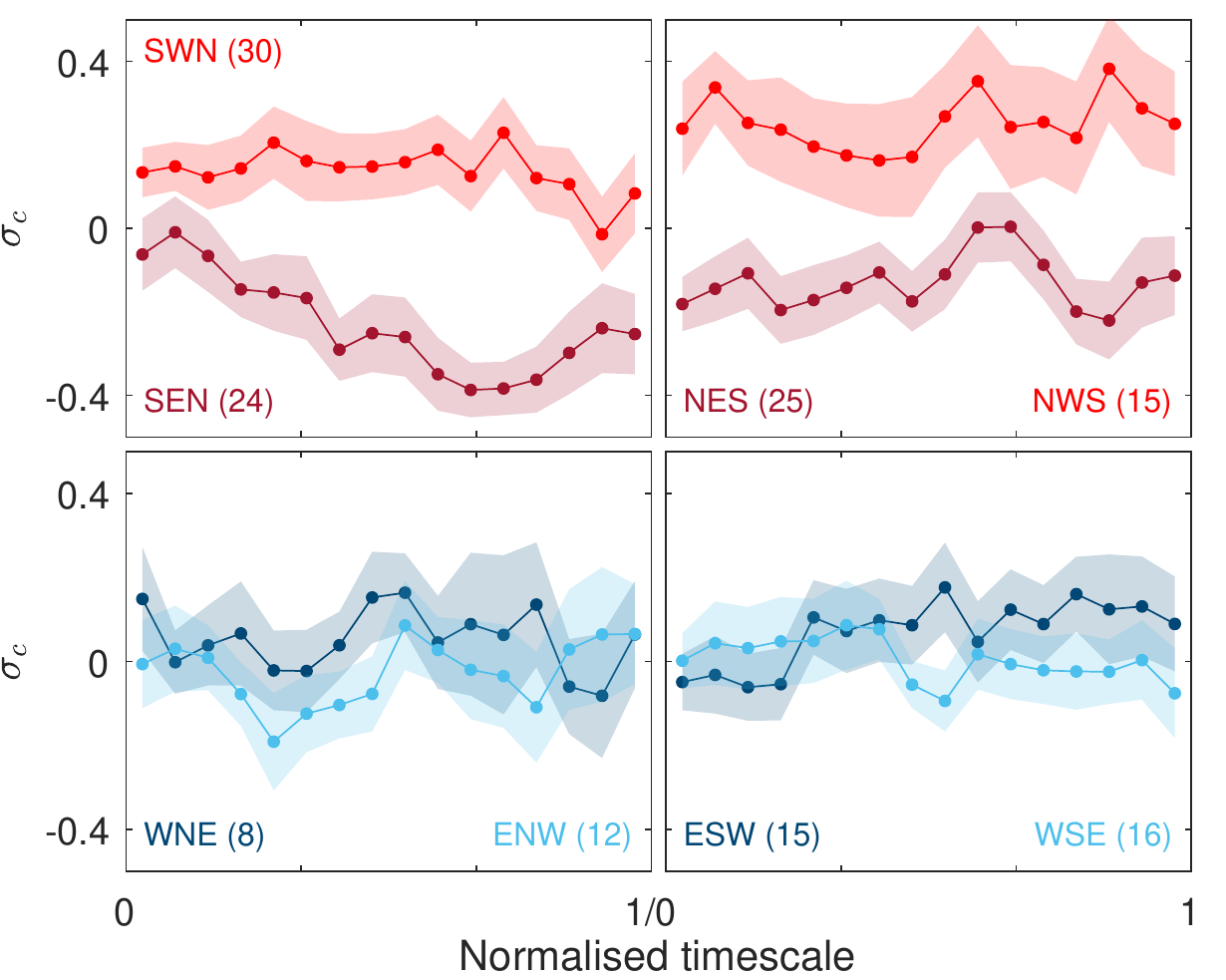}
    \caption{Superposed epoch profiles of $\sigma_\textrm{c}$ for flux ropes with a low inclination (red lines) and high inclination (blue lines) relative to the ecliptic plane, in a similar format to Figures~\ref{fig:SEA_figure1} and \ref{fig:SEA_figure2}.}
    \label{fig:SEA_figure3}
\end{figure}

Figure~\ref{fig:SEA_figure3} shows $\sigma_\textrm{c}$ profiles for flux ropes sorted into the eight categories devised by \citet{Bothmer98} and \citet{Mulligan98}. Each category has a three-letter designation, with the letters giving in sequential order the field polarity observed near the rope leading edge, central axis, and trailing edge. Polarities are given in terms of the solar cardinal directions, where N (north) $\simeq +B_z$, S (south) $\simeq -B_z$, E (east) $\simeq +B_y$, and W (west) $\simeq -B_y$. For example, the rope displayed in Figure~\ref{fig:event_example} is ENW-type, given the approximate rotation from east ($\theta = 0^{\circ},\phi = 90^{\circ}$) to north ($\theta = 90^{\circ}$) to west ($\theta = 0^{\circ},\theta = 270^{\circ}$). Four categories are associated with intrinsically left-handed ropes (SEN, NWS, ENW and WSE) and four with right-handed ropes (SWN, NES, WNE and ESW). Flux ropes were sorted into the eight categories using $H$, $\theta_0$ and $\phi_0$ from the Nieves-Chinchilla fits, with axis direction bins centred on each of the four cardinal directions and spanning $90^{\circ}$ in latitude and longitude. Ropes with axis directions closer to the $\pm x$ directions, which do not fall into the eight-category classification system described above, are generally more difficult to fit accurately with cylindrical models and have been excluded from the present analysis. Note that we have reverted to using unrectified $\sigma_\textrm{c}$ in this analysis, so that the mean field direction is inferred from the flux rope structure rather than the sector structuring previously used.

The top panels in Figure~\ref{fig:SEA_figure3} show profiles for ropes at a low inclination to the ecliptic (i.e., with eastward or westward axes), and the bottom panels show profiles for highly inclined ropes (i.e., with northward or southward axes). The profiles are paired in a way that matches south with south and north with north, and juxtaposes east and west. It can be seen that ropes with a westward axis (SWN and NWS) tend to have positive $\sigma_\textrm{c}$ and ropes with an eastward axis (SEN and NES) have negative $\sigma_\textrm{c}$. Given that the profiles show unrectified $\sigma_\textrm{c}$ values, this trend corresponds to a predominance of eastward-propagating Alfv\'enic fluctuations in all four low-inclination rope categories. Considering the sense of solar rotation, the eastward direction will tend to be anti-sunward at 1~au. However, this assessment is complicated if we envision flux ropes as closed loops with both ends connected to the Sun, i.e., both east and west are ultimately sunward on a closed loop. This issue is discussed further in Section~\ref{sec:disc}.

The four high-inclination profiles are more overlapping than the low-inclination profiles, with $\sigma_\textrm{c}$ values generally closer to zero. If the same east vs. west trend were seen as for the low-inclination ropes, the high-inclination profiles would be bimodal, i.e., W--E ropes would show a positive to negative change in $\sigma_\textrm{c}$, and vice versa; using solar wind sector phenomenology, high inclination ropes are mixed-sector intervals. However, there is little evidence of this bimodality in Figure~\ref{fig:SEA_figure3}, with only a very weak signature of it in the profiles shown in the bottom right panel. The relatively low number of events contributing to the high-inclination profiles may not be sufficient for any systematic bimodality to become distinguishable from the random statistical variations between events. There is also likely to be a dependence on the sign and magnitude of $y_0/R$ for the high inclination ropes, but the low event numbers preclude further investigation in terms of superposed epoch profiles. Alternatively, it may be that the east-west trend in $\sigma_\textrm{c}$ is only present when the east-west field in question forms part of the rope's axial field (top panels) rather than the rope's outer, poloidal field (bottom panels). The polarity of the axial field appears to dominate the overall $\sigma_\textrm{c}$ sign across the flux rope. Northward and southward field may be either sunward or anti-sunward, which may explain why there is no consistent north-south trend in $\sigma_\textrm{c}$ in any of these averaged profiles.

\section{Discussion}
\label{sec:disc}

\begin{figure}
	\includegraphics[width=0.99\columnwidth]{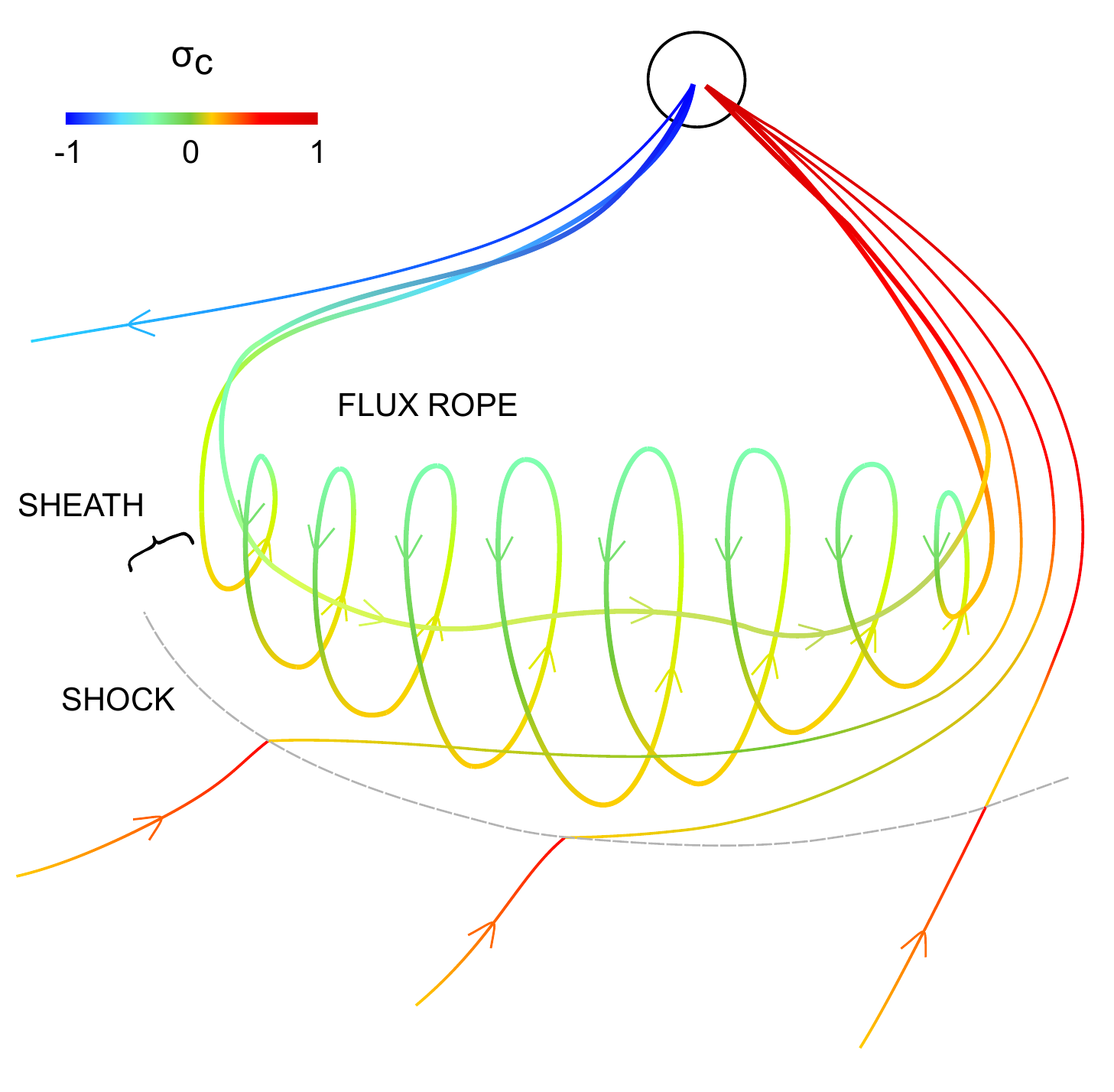}
    \caption{Cross helicity in an ICME with a flux rope, sheath and shock substructure, propagating through the Parker spiral. Solid lines show the mean field, with the cross helicity of Alfv\'enic fluctuations at smaller scales indicated by the line colour. Blues and reds show regions where fluctuations are propagating parallel and anti-parallel to the local field direction, respectively, with green shades indicating a balance of the two fluxes. Cross helicity values are low within the ICME, but the balance is tipped towards the anti-sunward direction throughout the ICME volume on average.}
    \label{fig:cartoon}
\end{figure}

\subsection{Solar origins}

Envisioned as closed loops with both footpoints anchored to the photosphere, ICME flux ropes could in principle carry a relatively balanced population of Alfv\'enic fluctuations from the corona and across the Alfv\'en critical point. Such a balanced state would arise if anti-sunward fluctuations have sufficient time (before the rising loop has crossed the critical point) to propagate up one side of the loop and down the other, where they would become sunward-directed. Mixing of fluctuations propagating in both directions around the loop could thus produce a balanced state right up to the critical point, which then becomes `frozen-in' and swept out with the supersonic flow at higher altitudes. This contrasts to open solar wind field lines, along which only an anti-sunward component is carried above the critical point. A number of sources have been suggested for the sunward fluctuations that are observed in the solar wind in interplanetary space, including reflection off the radial gradient in the Alfv\'en speed \citep[e.g.,][]{Chandran11}, parametric decay \citep[e.g.,][]{Bowen18} and generation within interaction regions \citep[e.g.,][]{Smith11}. In an ICME flux rope, in contrast, there is the intriguing possibility that the coronal source of anti-sunward fluctuations is also a source of the sunward fluctuations.

This closed-loop scenario was previously outlined by \citet{Good20}, and, in a similar vein, \citet{Borovsky19} have discussed how Alfv\'enic properties of solar wind at 1~au may be partly determined by the degree to which coronal source regions are open or closed. The low cross helicity of fluctuations in ICME flux ropes at 1~au may be a direct remnant of the coronal activity described above, or an indirect signature that develops via a turbulent cascade: balanced, low-frequency driving of the flux rope loop in the corona and beyond would produce balanced turbulence at the higher, inertial range frequencies that we have examined.

While the cross helicity magnitude is low, the balance is tipped towards the anti-sunward direction -- i.e., $\sigma_\textrm{c}^*$ is positive -- on average throughout the flux rope volume. Within the closed-loop framework described above, this could arise, for example, if balance does not typically extend along the full length of the flux rope loop at 1~au and is limited to some region centred on the loop apex. Thus, random spatial sampling of flux rope loops with balanced fluctuations near the apex and more imbalanced, anti-sunward fluctuations towards the legs would give a positive mean value of $\sigma_\textrm{c}^*$ overall. Figure~\ref{fig:cartoon} shows a schematic picture of an ICME with lower $|\sigma_\textrm{c}|$ around the rope apex and higher $|\sigma_\textrm{c}|$ in the rope legs connecting back to the Sun. The figure also shows higher $|\sigma_\textrm{c}|$ at higher $p$ (i.e. at larger crossing distances from the rope axis), consistent with Figure~\ref{fig:SEA_figure2}. Further analysis would be required to confirm whether ICME legs indeed have the higher $|\sigma_\textrm{c}|$ depicted in Figure~\ref{fig:cartoon}.

The extent to which counter-propagating fluctuations reach a balanced state would be dependent on a number of factors, including the time the flux rope spends in the corona during launch, the local Alfv\'en speed, and field line lengths. Global balance would most easily be achieved near the Sun, where field lines are shorter and the Alfv\'en speed is higher. The same factors are relevant when considering the large-scale incoherence of ICME structure: \citet{Owens20} finds that ICMEs are generally too large and expanding too rapidly at 1~au for information propagating at the Alfv\'en speed to travel from one longitudinal or latitudinal extremity of an ICME to the other.

A further complexity arises if we consider individual flux ropes as tubes that contain twisted field lines with varying lengths. For example, the widely used Lundquist flux rope model has longer, highly twisted field lines in the rope’s outer layers and shorter, less twisted field lines running through the core. Therefore, if field line length is a significant factor, longer field lines near the leading and trailing edges might be expected to have less balanced fluctuations in a Lundquist rope. Balance would be further reduced by Alfv\'en speeds tending to be lower near the rope edges, at least in magnetic clouds \citep{Owens20}. Cross helicity magnitudes are indeed relatively high towards the trailing edge (Figure~\ref{fig:SEA_figure1}), but not at the leading edge. In contrast, \citet{Good20} found higher $|\sigma_\textrm{c}|$ near both edges within a single magnetic cloud observed at 0.25~au; local interactions with the solar wind and opening of field lines by reconnection (further discussed below) were suggested in that work as potential causes of the higher cross helicity at the rope edges. Note that other rope models (e.g., Gold-Hoyle) have more uniform field line lengths and twists, and such ropes would be expected to have more uniform cross helicity across their radial profiles.

\subsection{Interplanetary interactions}

Another source of the increasingly more imbalanced, solar wind-like values of cross helicity towards the flux rope rear seen in Figures~\ref{fig:SEA_figure1} and \ref{fig:SEA_figure2} could be the erosion described by \citet{Dasso06}, whereby magnetic reconnection progressively peels away the outer field lines of the flux rope, often at the leading edge. In this scenario, the field lines remaining at the trailing edge, which were previously connected to the now-eroded leading edge field, form a connection to the open solar wind field and thus (along with the entrained plasma) acquire more solar wind-like properties. This process potentially occurs from the flux rope launch time onwards. \citet{Telloni20} recently reported how such erosion could explain significant changes in the large-scale MHD structure of an ICME flux rope observed by aligned spacecraft at 1 and 5.4~au, with the reconnection in this case occurring with a second, trailing ICME. 

Cross helicity in ICME sheaths at 1~au is likewise lower than in the solar wind on average, but not as low as in the flux ropes. Since sheaths at 1~au consist of solar wind that has gradually accumulated with ICME propagation in interplanetary space, an interplanetary origin of their reduced cross helicity (e.g., velocity shear) is perhaps more likely. While at low $|\sigma_\textrm{c}|$, anti-sunward fluctuations still predominate in sheaths: this is consistent with an admixture of pre-existing anti-sunward fluctuations swept up from the solar wind and a balanced, locally generated population that acts to reduce the overall $|\sigma_\textrm{c}|$. The same effect has been identified in fast--slow solar wind stream interaction regions \citep[SIRs; e.g.,][]{Smith11}, which are in many respects analogous to ICME sheaths. 

In the preceding discussion, it has been suggested that the low cross helicity in ICME flux ropes is an intrinsic property that originates in the corona, with the rising $\sigma_\textrm{c}^*$ through the flux rope profile being due to the effects of reconnection or variable field line lengths. An alternative interpretation is that the entire sheath-ICME complex behaves like an SIR, with the low cross helicity primarily arising in interplanetary space via the ICME-solar wind interaction. Some evidence that supports this hypothesis is found in the bottom two panels of Figure~\ref{fig:SEA_figure2}. Taking the presence of shocks as an indicator of stronger ICME-solar wind interaction, it can be seen that ICME flux ropes driving shocks have enhanced levels of $E^*_\pm$ and lower $\sigma_\textrm{c}^*$ relative to ropes that do not drive shocks. A possible interpretation of Figures~\ref{fig:SEA_figure1} and Figures~\ref{fig:SEA_figure2} is that $E^*_\pm$ are both amplified at the sheath leading edge and then decay back to ambient solar wind levels \citep[e.g.,][]{Pitna17} with distance behind the leading edge. Since the ambient $E^*_-$ is lower than the ambient $E^*_+$, this could explain why the drop in $E^*_-$ through the ICME is more pronounced. However, it is not clear what physical mechanism could produce such behaviour. The shock (or sheath leading-edge wave, in the case of sheaths without shocks) cannot be the direct cause of the low cross helicity within the flux rope, since the plasma in the flux rope has not at any stage passed through the shock. Likewise, velocity shears are unlikely to be a cause, since they occur very infrequently or weakly within ICME plasma \citep{Borovsky19}. However, velocity shear may be the cause of the $E^*_\pm$ enhancement centred at the sheath-flux rope interface.

Coronal and interplanetary modelling would be required to test the range of hypotheses discussed in this section. Besides modelling, additional measurements of ICME cross helicity closer to the Sun would bring some clarity. PSP and \textit{Solar Orbiter} \citep{Muller20} are now regularly observing ICMEs in the inner heliosphere \citep[e.g.,][]{Davies21,Weiss21,Palmerio21,Mostl22}. As the number of ICMEs observed by these spacecraft continues to grow, a statistical analysis of ICMEs similar to the one presented here will be possible for sub-1~au heliocentric distances.

\section{Conclusion}

In this study, low cross helicity of fluctuations at inertial range scales has been identified as a typical property of ICME plasma at 1~au. Low cross helicity is indicative of a balance between Alfv\'enic fluxes propagating parallel and anti-parallel to the background magnetic field. The low cross helicity in ICMEs is evident from average values across intervals (Section~\ref{sec:mean}) and from superposed epoch analyses (Section~\ref{sec:ICME_dependence}), the latter revealing a systematic variation through the flux rope and sheath substructures: the sheath-flux rope complex represents a local depression in cross helicity embedded in the solar wind flow. We suggest that low cross helicity should be considered in the same light as more established interplanetary signatures of ICMEs \citep{Zurbuchen06}. The relatively low cross helicity in ICME flux ropes may primarily originate from their closed field structure, in contrast to the higher cross helicities and more open field structure of the solar wind.

\section*{Acknowledgements}

We thank the \textit{Wind} instrument teams for the data used in this study. SWG is supported by Academy of Finland Fellowship grants 338486 and 346612 (INERTUM), and Project grant 310445 (SMASH). EKJK, MA-L and JES are supported by funding from the European Research Council under the European Union’s Horizon 2020 research and innovation programme, grant 724391 (SolMAG). This work was performed within the framework of Academy of Finland Centre of Excellence grant 312390 (FORESAIL).

\section*{Data Availability}

Data were obtained from the publicly accessible CDAWeb archive at https://cdaweb.sci.gsfc.nasa.gov. The ICME catalogue used in this study is found at https://wind.nasa.gov/ICMEindex.php.




\bibliographystyle{mnras}
\bibliography{references} 








\bsp	
\label{lastpage}
\end{document}